\documentclass[runningheads]{llncs}
\usepackage[T1]{fontenc}
\usepackage{graphicx}
\usepackage{amsmath}
\begin{document}
\title{Metastable states are precursors of Turing patterns in Adaptive Networks}
\titlerunning{Metastable States in Adaptive Networks}
%
\author{S. Nirmala Jenifer\inst{1,2}\orcidID{0009-0009-7389-0116} \and
Timoteo Carletti\inst{1}\orcidID{0000-0003-2596-4503}} 
\authorrunning{Jenifer et al.}
%

\institute{Department of Mathematics and naXys, Namur Institute for Complex Systems,
University of Namur, Namur, Belgium \and 
Department of Physics, Bharathidasan University, Tiruchirappalli 620024, Tamil Nadu, India \\
\email{\{nirmala-jenifer.selvaraj,timoteo.carletti\}@unamur.be}}


%
\maketitle              
\begin{abstract}

In networked reaction-diffusion systems,  emergence of patterns from a homogeneous equilibrium can be described by Turing instability. Recently, Turing theory has been developed to include adaptive networks where, i.e., link weights change as a function of the states of the nodes. In this work we show that metastable states, defined as long living transient epochs with unique dynamical properties, could be observed in a range of parameters between Turing patterns and absence of any spatial pattern. Stated differently in the metastable regime the system ``switches'' between the homogeneous solution and the Turing pattern, never settling in any of those states, i.e., alternating silent and spiking regimes. By using phase diagrams and recurrence plots, we have numerically shown the existence of metastable states for the Brusselator model. Moreover, the dependence of the emergence of the metastable states on the adaptive control parameter has been characterized. The applicability of methods and ideas proposed in this work go beyond the considered case, they can help us in understanding  conditions under which metastable states arise and  could thus be informative to predict and control the latter.
 

\keywords{Adaptation \and Metastable \and complex networks \and Turing Patterns.}
\end{abstract}
%
%
%


%


\section{Introduction}
\begin{figure}
	\centering
	\includegraphics[width=0.69\linewidth]{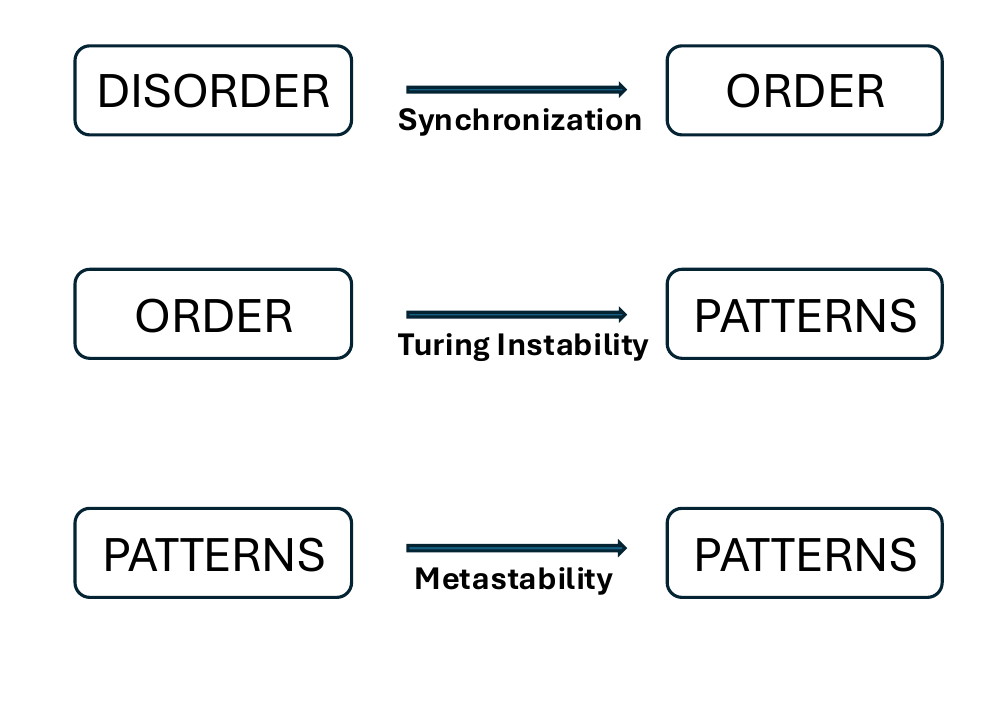}
	
	\label{fig:metastability}
    \caption{\textbf{Possible mechanisms of transition among relevant dynamical system states.}}
\end{figure}

Complex systems are around us; from social networks, power-grids, internet, financial networks~\cite{Kwapien2012} to  human brain~\cite{bullmore2009complex}. Their main signature is that they exhibit a global behavior, i.e., as a whole, different from the behaviors of the individual units that comprise the system~\cite{Boccaletti2006}.
Such behavior often emerges as patterns, where the interaction between the units give raise to collective dynamics unpredictable from the behaviors of the individual units. To model those systems, we use networks where units are represented by nodes and interactions between the latter are represented by links~\cite{Kuramoto1975,Iacopini2019}. Hence, the emergent behaviors can be modeled by using dynamical systems defined on complex networks. These emergent behaviors or patterns, very often mark a transition from disorder to order~\cite{gambuzza2021stability}; they can arise in systems, where nodes with an initial irregular dynamics adjust their behaviors due to the interaction between them and determine a resulting coherent dynamics, e.g., synchronization~\cite{jenifer2024synchronizability} (see top panel of Fig.~\ref{fig:metastability}).

On the other hand, it has been shown that the patterns can emerge from the otherwise ordered or homogeneous states~\cite{turing1990chemical} (see middle panel of Fig.~\ref{fig:metastability}). As an example, in reaction-diffusion systems where reactions occur in nodes and links provide paths of species diffusion, the latter process can destabilize the homogeneous solution and resulting into a new stable patchy solutions. This phenomenon has been modeled and explained by Alan Turing~\cite{turing1990chemical} inspired by the presence of many regular spatio-temporal behaviors, such as spots and stripes on animal skins or oscillations in chemical reactions~\cite{Nicolis1977}. Those models involve reaction, i.e., creation and destruction of basic quantities, e.g., species, and movement of the latter in space, i.e., diffusion. In several real applications, one can assume the substrate upon which diffusion occurs to be discrete, i.e., a network, there reactions occur in the nodes and diffusion across the links connecting pairs of nodes. Turing instability is rooted on the fact that the existing stable homogeneous solution becomes unstable because of the diffusive process, leading to patterns. The emergence of Turing patterns in networks has been first studied in~\cite{nakao2010turing}. Recently, Turing patterns were studied in adaptive networks~\cite{dorchain2025turingpatternsadaptivenetworks} where, i.e., the network structure (weights of the links) evolves as a consequence of the node dynamics~\cite{gross2007adaptive,berner2023adaptive}. We assign weights~\cite{kumpula2007emergence} to links to differentiate their strength and moreover the latter adapt as a function of the nodes state. Such a coevolution creates a feedback loop, where the state of the nodes affects the connectivity of the network and the evolving link weights influence the node dynamics~\cite{GB2008,berner2023adaptive}. Adaptions play a crucial role while  systems go through transitions from one type of synchronization to other~\cite{nirmala2024synchronization} and to induce synchronization~\cite{PhysRevE.110.064305}. Adaptations in reaction-diffusion system also promotes Turing patterns from homogeneous solutions and helps to sustain Turing patterns in small networks resulting from pruning links in larger networks~\cite{dorchain2025turingpatternsadaptivenetworks}. 

Let us observe that many organisms develop patterns passing from an already heterogeneous state to another, not necessarily starting from the homogeneous condition~\cite{kelso1995dynamic} (see bottom panel of Fig.~\ref{fig:metastability}). These changes are generally related to transitions in the  metastable states. The latter are interesting and widespread; they are characterized by long living and transient ``shifts'' where the system seems to approach an equilibrium to then suddenly going away from it, moving back and forth toward one another in a never ending way~\cite{beim2019metastable}. They are mainly observed as transitions between different neuronal states such as spikes and silences in neuronal models \cite{hindmarsh1984model}, transitions to different modes of oscillations in the brain \cite{michel2018eeg} and the emergence of seizure from local synchronization of
neuronal ensembles \cite{wenzel2019acute}. Finally, they have also been observed in a wide range of real and synthetic systems~\cite{brinkman2022metastable,roberts2019metastable}.

 The goal of this work is to show that metastable states precede the emergence of Turing patterns in adaptive networks which manifest as bursty behaviors~\cite{kelso2012multistability}. In the latter the system seems to first move toward the homogeneous equilibrium, but then suddenly it goes away on the way to the Turing patterns; at this point adaptation enters into play and the changes in link weights push the system back to the homogeneous equilibrium. Those shifts repeat indefinitely for a given value of a control parameter. We claim that for intermediate values of this control parameter, the system exhibits metastable states, while if the values are large, the system settles into Turing patterns or wave. If the control parameter vanishes, then the system will remain in the homogeneous state. Our claim will be supported by dedicated numerical simulations to prove that the observed behavior can indeed be ascribed to a metastable state. As we increase the value of the control parameter starting from zero, the system stays close to the initial homogeneous solution for shorter and shorter times, then it exhibits an oscillatory behavior to eventually splits into a new heterogeneous stationary solution. Our work has thus two goals, first to provide a numerical evidence for the existence of metastable states and second to show how these states disappear and Turing patterns emerge in the presence of adaptations. 
 
 In the rest of the manuscript, in Section \ref{sec:The model}, we define the Brusselator model, in Section \ref{sec:Metrics of metastable states}, we provide some indicators of metastable states mainly from time series analysis and recurrence analysis, and the transitions to the Turing patterns from the meta stable states and back. In Section \ref{sec:Conclusion} we conclude our discussion. 
 
\section{The Model}
\label{sec:The model}

We consider a globally connected, weighted network consisting of nodes hosting reacting species whose densities are denoted by $u_i$ and $v_i$; the latter diffuse through the available links and the rate of diffusion is controlled by the positive link weights, as well as by the model parameters. We consider Brusselator model~\cite{PrigogineNicolis1967}, a stereotype example of system exhibiting Turing patterns to show the emergence of metastable states. The evolution of the Brusselator state variables $u_i$ and $v_i$ can be described by Eqs.~\eqref{eq:modeluvwBrusu} and~\eqref{eq:modeluvwBrusv}
\begin{eqnarray} 
 \frac{du_i}{dt}  &=& 1 - (b+1)u_i + c u^2_{i}v_i +D_u\sum_j L_{ij}(t) u_j \label{eq:modeluvwBrusu}\\
    \frac{dv_i}{dt} &=& bu_i - cu^2_{i}v_{i}+D_v\sum_j L_{ij}(t) v_j \label{eq:modeluvwBrusv}\\
   \frac{dw_{ij}}{dt} &=& a_{ij} \left(A^*-A_{\mathit{loc}}(u_i,u_j,v_i,v_j)\right) k(w_{ij})\label{eq:modeluvwBrusw}\, ,
\end{eqnarray} 
where $b>1$ and $c>0$ are model parameters, the equilibrium of the reaction part is given by $u^*=1$ and $v^*=b/c$. $D_u>0$, resp. $D_v>0$, is the diffusion coefficient of species $u$, resp. $v$. Eq.~\eqref{eq:modeluvwBrusw} describes the response function, i.e., how link weights, $w_{ij}$, adapt in function of the system state, $u$ and $v$; $A_{\mathit{loc}} (u_i,u_j,v_i,v_j)$ is a measure of the local distance of the nodes densities $u$ and $v$ with respect to the equilibrium point:
\begin{equation}
\label{eq:modelw3a}
A_{\mathit{loc}} (u_i,u_j,v_i,v_j)= \frac{(u_i(t)-u^*)^2+(u_j(t)-u^*)^2+(v_i(t)-v^*)^2+(v_j(t)-v^*)^2}{4}\, .
\end{equation}
\begin{figure}
    \centering
    \includegraphics[width=0.85\linewidth]{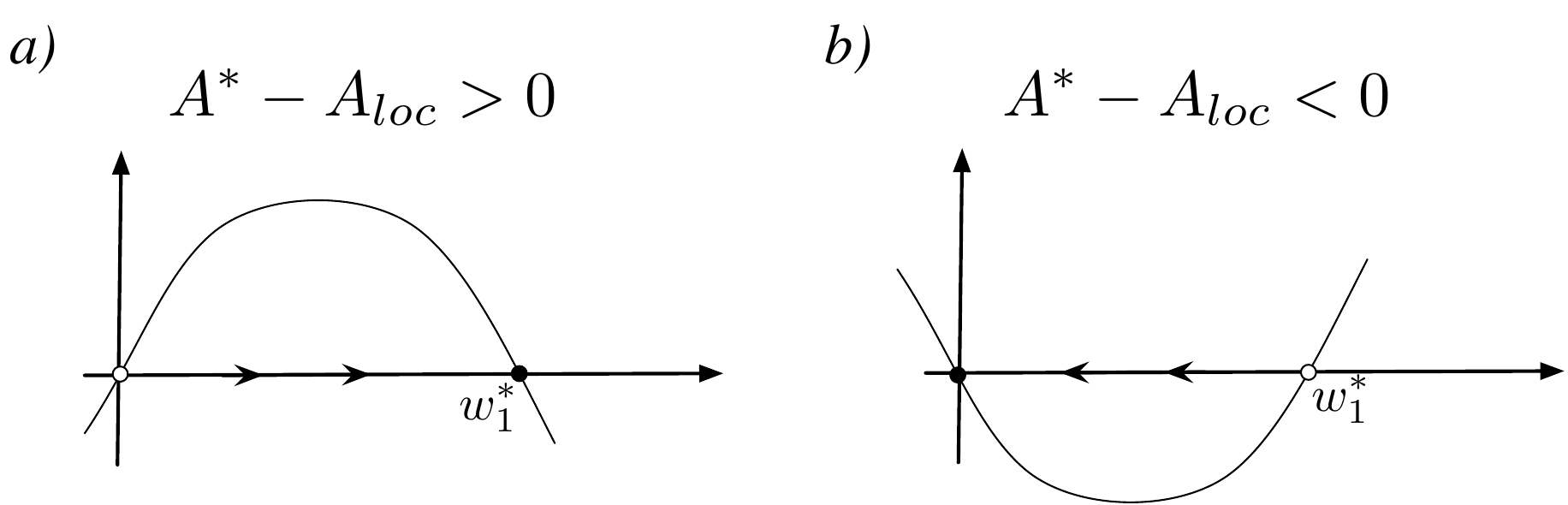}
    \caption{\textbf{Cartoon representing generic behaviors for the nonlinear adaptive response functions for the link weights evolution}. a) response function with negative concavity, $w_1^*$ is a stable fixed point (black circle); panel b) response function with positive concavity, $w_1^*$ is an unstable fixed point (white circle). $A_{\mathit{loc}}$ is the local patterns amplitude and the arrows on the horizontal axis denote the time evolution of weights $w_{ij}(t)$ according to Eq.~\eqref{eq:modeluvwBrusw}, i.e., the arrows point in the direction of increasing weights.}
    \label{fig:responsecurve}
\end{figure}
Finally, $k(w)=w(w_1^*-w)$, with $w_1^*=1/2$, is the chosen adaptive function depending on the control parameter $A^*\geq 0$. Let us observe that the latter function recalls the Hebbian, or anti-Hebbian, learning rule used in neuroscience~\cite{AA2009,AA2011}. Fig.~\ref{fig:responsecurve} represents the response function with negative (a) and positive (b) curvature and provides the conditions under which the system moves towards or away from the equilibrium weight, $w_1^*$. In the following we fix $b = 4.0$, $c = 6.5$, $D_u = 0.07$ and $D_v = 0.7$. We also fix the size of the network to $n=20$ nodes and we will be interested in the role of the parameter $A^*$. The initial conditions for $u$, resp. $v$, are drawn from the uniform distribution $U[-\delta,\delta]$ centered in $u^*$, resp. $v^*$; the initial weights are distributed uniformly at random in the interval $(w_1^*-\delta,w_1^*)$ and we set $\delta=0.01$. Because of our choice of initial conditions $A_{loc}(0)\sim \delta^2$, this will hence set a ``natural'' scale for $A^*$~\cite{dorchain2025turingpatternsadaptivenetworks}. We numerically integrate the systems to obtain the time series for several values of $A^*$, we further analyze these to provide the numerical evidence for the existence of the metastable states and the system transitions from the stable homogeneous equilibrium to new heterogeneous equilibria passing through metastable states.

\section{Metrics of metastable states}
\label{sec:Metrics of metastable states}

\begin{figure}
	\centering
	\includegraphics[width=0.95\linewidth]{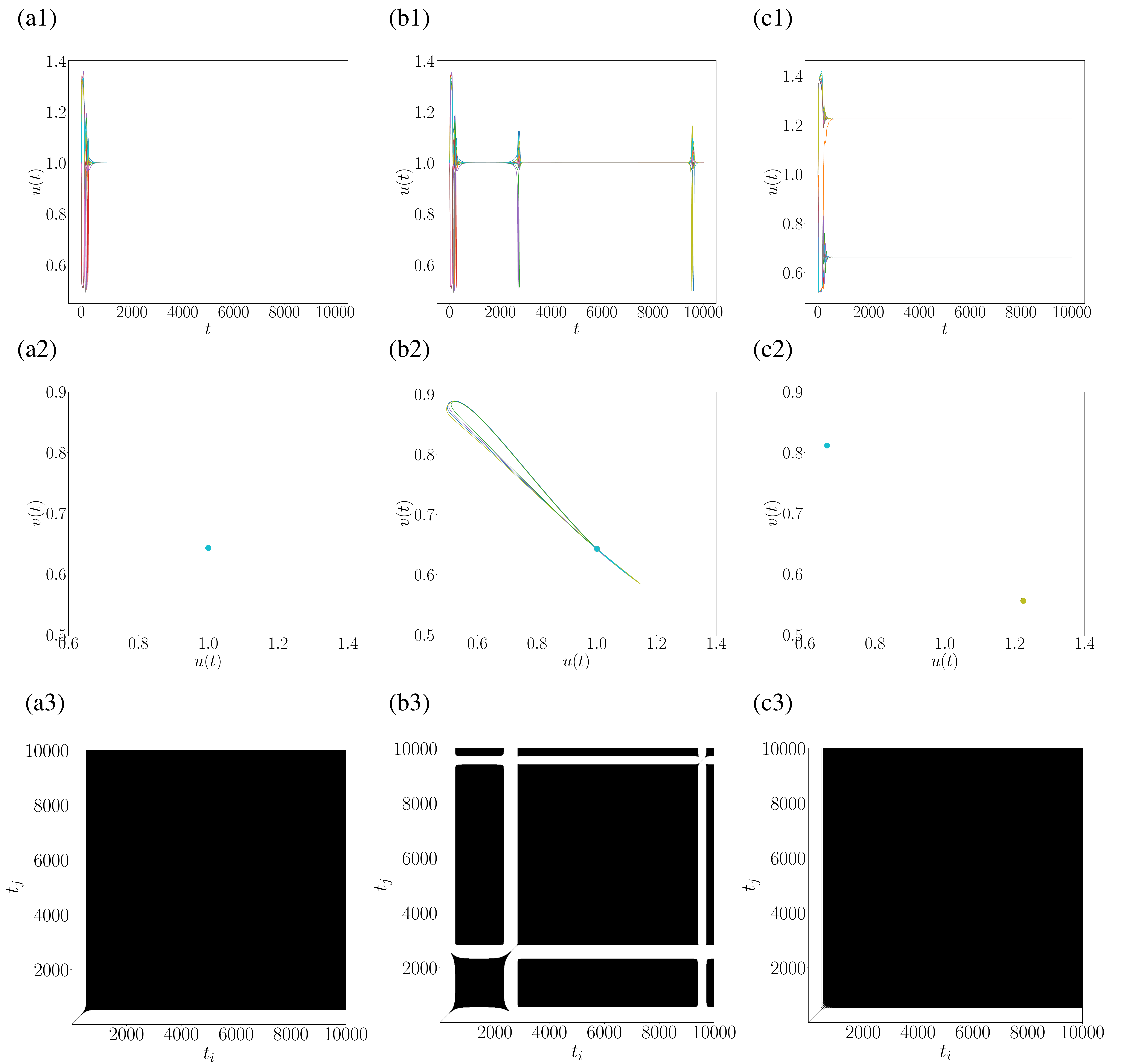}
	
	\caption{\textbf{Existence of metastable states}. Dynamical behaviors for $A^*=0.0$ (left column) correspond to homogeneous equilibrium, $A^*=0.0001$ (middle column) metastable states, $A^*=0.05$ (right column) Turing instabilities. Top row. Time evolution of $u_i(t)$. Middle row. State plot $u_i(t)$ vs $v(_it)$ showing fixed point (a2), metastable state switching between a fixed point and oscillations (b2) and Turing patters (c2). Bottom row. Recurrence plots: fixed point solution (a3), metastable states (b3) and Turing patters (c3). Here, black points represent elements for which $R_{ij} = 1$ while white points correspond to $R_{ij} = 0$.}
	\label{fig:figure}
\end{figure}

To identify the presence of metastable states, several methods have been proposed. The obvious and primary one is to plot the time-series (see top panels of Fig.~\ref{fig:figure}); if in the latter we observe more than one dynamical behavior, then we have a first numerical evidence of metastable states. Then, to identify different dynamical regimes, we display the state plots $u(t)$ vs $v(t)$; this will provide us with clear different dynamical regimes of metastable states (see middle panels of Fig.~\ref{fig:figure}). To be more precise, we plot the recurrence plots which provides us information about the similarities, or lack thereof, between the values of $u_k(t)$ at different times (see bottom panels of Fig.~\ref{fig:figure}). Formally, the recurrence plot is obtained by computing 
\begin{equation}
	R_{ij} = \Theta(\epsilon - D_{ij})\, ,
\end{equation}
where $\Theta$ is the Heaviside step function, $\epsilon>0$ and
\begin{equation}
	D_{ij} = \sqrt{\sum _{k=1 }^n (x_k(t_i) - x_k(t_j))^2}\, ,
\end{equation}
the Euclidean distance between the orbit of species $k$ computed at times $t_i$ and $t_j$ (we hereby assume to discretize the time interval). Hence $R_{ij} = 0$ if $D_{ij} \ge \epsilon $ and $R_{ij} = 1$ if $ D_{ij} < \epsilon $. We can clearly observe different dynamical regimes, homogeneous solution (left bottom panel of Fig.~\ref{fig:figure}) and Turing patterns (right bottom panel of Fig.~\ref{fig:figure}), they are characterized by a coherence region $R_{ij} = 1$ (black) after a short transient $R_{ij} = 0$ (white); while in the middle panel of Fig.~\ref{fig:figure}, we see two regions (displayed in black) where  $R_{ij} = 1$ separated by ``stripes'' corresponding to $R_{ij} = 0$ (white). This is a signature of metastable states. 
For the given adaptation function for weights, Eq.~\eqref{eq:modeluvwBrusw}, when $A^* = 0$, the system remains in the homogeneous equilibrium, while for larger values of $A^*$, the homogeneous solution becomes unstable determining Turing patters. However, for intermediate values of $A^*$, we have observed ``bursty'' patterns characterized by quick ``firing of spikes'' followed by quiescent period~\cite{dorchain2025turingpatternsadaptivenetworks,PhysRevE.111.021001}. These patterns resemble metastable states emerging from multi-stable states \cite{kelso2012multistability}, since we observe periods of oscillations followed by silence, i.e., orbits very close to the homogeneous solution. A similar behavior has been previously observed in Hindmarsh-Rose neurons where an adaptation current causes the switching between the spiking and the silence state~\cite{hindmarsh1984model}.



To perform our analysis, we first obtain the time series over a long enough time interval, i.e., $[0,10000]$, by using parameters fixed to the values used in~\cite{dorchain2025turingpatternsadaptivenetworks}. The {\em time series} for all $u_k(t)$ are displayed in top panels of Fig.~\ref{fig:figure}; the left panel corresponds to $A^*=0$ and we can observe the convergence of $u_k(t)$ to the equilibrium $u^*=1$. The right panel, $A^*=0.05$, shows the existence of Turing patterns, while in the middle panel, $A^*=0.0001$, we observe the above claimed bursty behavior. To achieve a better understanding of the dynamics, we have then displayed the {\em state plot} $u_k$ versus $v_k$ in middle panels of Fig.~\ref{fig:figure}, by using orbits shown in top panels; in the left panel we can observe a fixed point while the right panel shows two points, signature of Turing pattern. The middle panel clearly exhibits an oscillatory behavior where the system switches between the two points shown in the right panel, this is the signature of metastable states. Eventually, in the bottom panels of Fig.~\ref{fig:figure}, we have plotted the {\em recurrence plot}.

\begin{figure}
    \centering
    \includegraphics[width=0.60\linewidth]{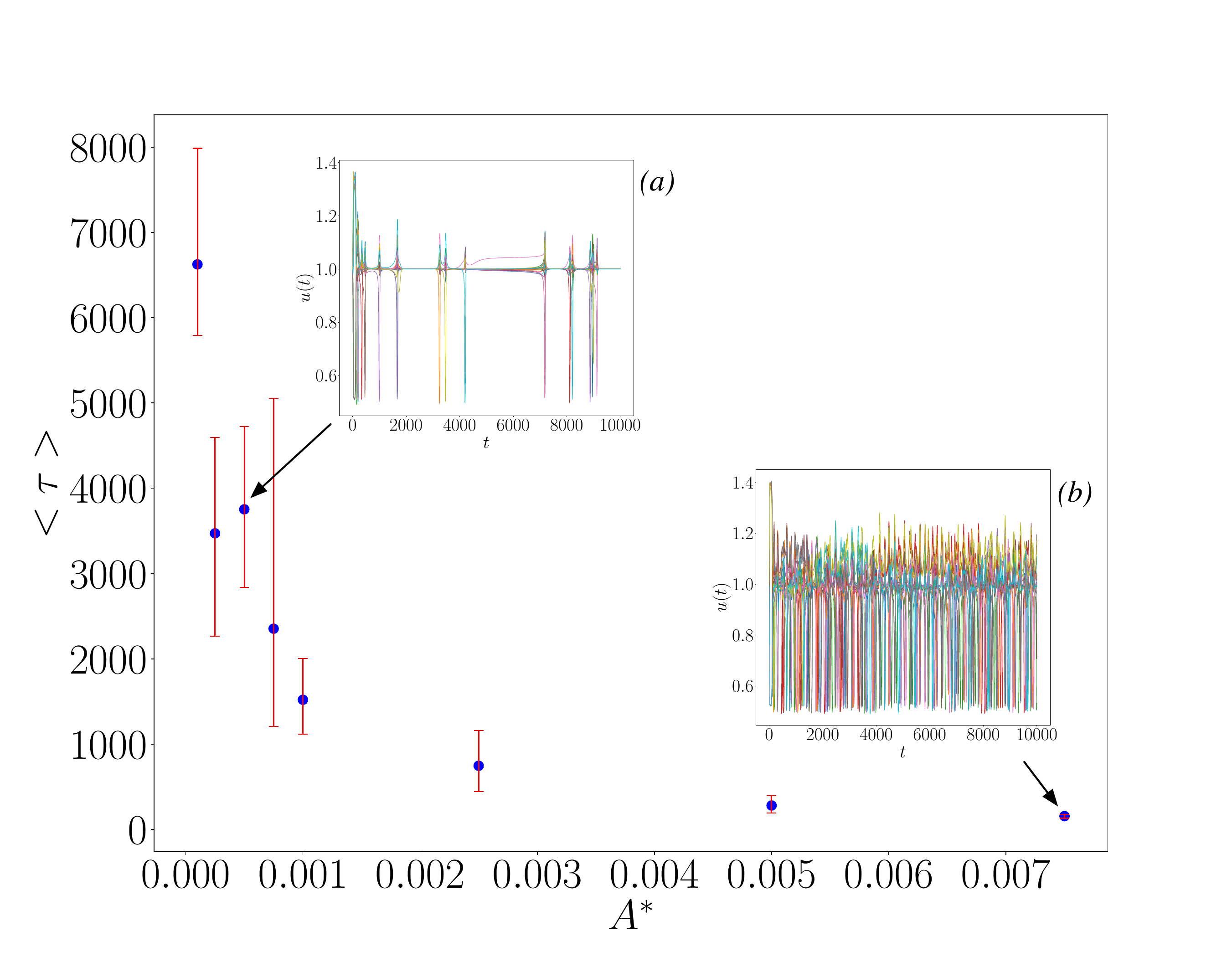}
    \caption{\textbf{Average of longest silent interval as a function of $A^*$}. We can observe that as $A^*$ increases, $\langle \tau\rangle$ decreases (blue dots) indicating that the system spends lesser and lesser time close to the homogeneous solutions $u^*$. At $A^*\sim 0.0075$, $<\tau> \sim 0$ showing that the system no longer returns to the original homogeneous solution. Error bars, shown in red, denote the maximum and minimum of the longest duration of the silent phases. The insets (a) and (b) correspond to the time series for $A^* = 0.0005$ and  $A^* = 0.0075$.}
    \label{fig:tauvsA}
\end{figure}

Finally, we study the way the system transits from the homogeneous state to the Turing patterns passing through the metastable state. For intermediate values of $A^*$ we have observed the bursty behavior characterized by silent phases of duration $t^*$ followed by sudden spikes; our goal is to study the dependence of the longest duration on $A^*$. For a given value of $A^*$, we numerically integrate the system $10$ times and for each one we compute the longest duration of the silent phase, $\tau^*_i$, $i=1,\dots,10$. More precisely the silent duration is defined as the longest interval of time for which $|u_k(t)-u_*|< \theta$, with $\theta = 0.01$. Eventually we defined $\langle \tau\rangle  = \sum_i  \tau^*_i/10$. To appreciate the variations about the average we also compute the minimum and maximum of these values, $\min \tau_i^*$ and $\max \tau_i^*$. The results are shown in Fig.~\ref{fig:tauvsA} where we plot $\langle \tau\rangle$ (blue points) for $A^* \in\{  0.0001,0.00025,0.0005,0.00075,0.001,0.0025,0.005,0.0075\}$, the vertical red bars denote the interval $[\min \tau_i^*,\max \tau_i^*]$. For small positive $A^*$, the silent phases can be very long and $<\tau>$ is large (see Fig.~\ref{fig:tauvsA} inset (a)). We can observe that as we increase $A^*$, the maximum duration of time that the system stays close enough to $u^*$ decreases; at $A^*\sim  0.0075$ we get $<\tau> \sim 0$, namely the system starts oscillate more often (see Fig.~\ref{fig:tauvsA} inset (b)). For larger values of $A^*$, the system no longer returns to the original homogeneous solution and moves toward a Turing pattern, stationary or wave (data not shown).

\section{Conclusion}
\label{sec:Conclusion}
In dynamical systems, the more relevant yet underexplored states are metastable states which occur when the system jumps between different patterns of activities. There could be several reasons to induce these transitions and one of them is adaptations. Indeed, J.L. Hindmarsh and R. M. Rose showed that the system oscillates between silence and bursts caused by adaptation currents~\cite{hindmarsh1984model}. In reaction-diffusion systems we observe metastable states manifested as a continuous and bursty switching between the homogeneous solution and the patchy one. The system exhibits metastable states for the range of control parameters before the system transits to Turing patterns, being the latter stationary or oscillating. By using the Brusselator model we have numerically shown that in adaptive networks, metastable states occur for the values of the adaptive control parameter in between those returning stable homogeneous equilibrium and Turing patterns. We have used phase diagrams and recurrence plots to numerically prove the existence of the metastable states. We have computed the longest duration in time that the system remains sufficiently close to the homogeneous steady state as we vary the adaptive control parameter $A^*$. For small values, the system exhibits very long silent phase, i.e., it remains close enough to the homogeneous equilibrium. Then we observed that as we increase $A^*$, the duration the system stays close to $u^*$ decreases and more often exhibits oscillating behavior. If we increase $A^*$ even more, the system moves toward Turing patterns. 

Our results would help us to understand adaptive networks and bring us closer to predict and control the emergence of metastable states in real-world systems with adaptations, for instance the transitions between different brain states as during epilepsy.

%
%
%
\section*{ACKNOWLEDGMENTS}

We thank Maxime Lucas for his valuable discussions. The work of S.N.J. is supported by the Institutional call for doctoral fellowship (CERUNA), University of Namur.
 \bibliographystyle{splncs04}
 \bibliography{reference}

\end{document}